\documentclass[a4paper,twocolumn, nofootinbib,superscriptaddress,aps,prl, 14pt,eqsecnum,notitlepage,showkeys]{revtex4-1}
\usepackage{multirow}
\usepackage{graphicx}
\usepackage{bm}
\usepackage{dcolumn}
\usepackage{hyperref}
\usepackage{gensymb}
\usepackage{amsmath}
\usepackage{dcolumn}    % Align table columns on decimal point
\usepackage{ifpdf}
\usepackage{amssymb,lineno,amsfonts}
\usepackage{graphicx}   % Include figure files
\usepackage{bm}         % bold math
\usepackage{bbm}
\usepackage{mathrsfs}
\usepackage{upgreek}
\usepackage{mathtools}
\usepackage{epstopdf}
\usepackage{setspace}
\usepackage{hyperref}
\usepackage{natbib}
\usepackage{esvect}
\usepackage{amsmath}
\usepackage[usenames,dvipsnames]{xcolor}
\definecolor{med-blue}{RGB}{25,25,112}
\hypersetup{colorlinks, linkcolor={blue},citecolor={blue}, urlcolor={blue}}
\usepackage{times }
\usepackage [english]{babel}
\usepackage [autostyle, english = american]{csquotes}
\MakeOuterQuote{"}
\begin{document}
	\title{Temperature dependence of the Spin Seebeck effect in a mixed valent manganite}
	\author{Avirup De}
	\author{Arup Ghosh}
	\author{Rajesh Mandal}
	\affiliation{Department of Physics, Indian Institute of Science Education and Research}
	\author{Satishchandra Ogale}
	\author{Sunil Nair}
	\affiliation{Department of Physics, Indian Institute of Science Education and Research}
	\affiliation{Centre for Energy Science, Indian Institute of Science Education and Research,\\ Dr. Homi Bhabha Road, Pune, Maharashtra-411008, India}
	\date{\today}
	\begin{abstract} 
		We report on temperature dependent measurements of the Longitudinal Spin Seebeck Effect (LSSE) in the mixed valent manganite La$_{0.7}$Ca$_{0.3}$MnO$_3$. By disentangling the contribution arising due to the Anisotropic Nernst effect, we observe that these two thermally driven phenomena vary disparately with temperature. In a narrow low temperature regime, the LSSE exhibits a $T^{0.55}$ dependence, which matches well with that predicted by the magnon-driven spin current model. Across the double exchange driven paramagnetic-ferromagnetic transition, the LSSE exponent is significantly higher than the magnetization one. These observations highlights the importance of individually ascertaining the temperature evolution of different mechanisms which contribute to the measured spin Seebeck signal.
	\end{abstract}        
	% insert suggested PACS numbers in braces on next line
	\pacs{Pacs}
	% insert suggested keywords - APS authors don't need to do this
	\maketitle
The Spin Seebeck Effect (SSE) \cite{uchida2010spin,uchida2008observation,uchida2010observation,kikkawa2013longitudinal, uchida2014quantitative, giles2017thermally} pertains to the generation of a thermally induced magnonic spin current in a magnetic material subjected to a temperature gradient. The favored means of measuring this spin current is in the form of the longitudinal Spin Seebeck Effect (LSSE), where a normal metal (NM) with a large spin orbit coupling (typically Pt) is deposited on the magnetic material, and a temperature gradient is applied perpendicular to the plane of the interface \cite{meier2015longitudinal}. The inverse Spin Hall Effect (ISHE) enables the conversion of the spin current to a measurable electrical potential difference $E_{ISHE} \propto \vec{j}_S \times \vec{\sigma} $ (where $\vec{J}_S$ is the spin current density and $\vec{\sigma}$ is the spin polarization of the itinerant electrons in the NM layer) \cite{saitoh2006conversion}. Junctions comprising of the ferrimagnetic insulator Y${_3}$Fe${_5}$O${_{12}}$ (YIG) and Pt are now recognized as being model systems for investigations of the LSSE, and recent measurements have focused on the dependence of the LSSE signals on control parameters like temperature \cite{uchida2014quantitative,iguchi2017concomitant,jin2015effect,wang2015spin}, magnetic field \cite{kikkawa2015critical, ritzmann2015magnetic} and sample thickness \cite{kehlberger2015length,guo2016influence}. Temperature dependent LSSE measurements have also revealed a remarkable correlation of this quantity with intrinsic properties of the magnetic material, and a few strongly correlated systems have been explored in this fashion \cite{ramos2013observation, geprags2016origin, wu2017longitudinal}. 

\begin{figure}
	\centering
	\vspace{-0.1cm}
	\hspace{-0.2cm}
	\includegraphics[scale= 0.32]{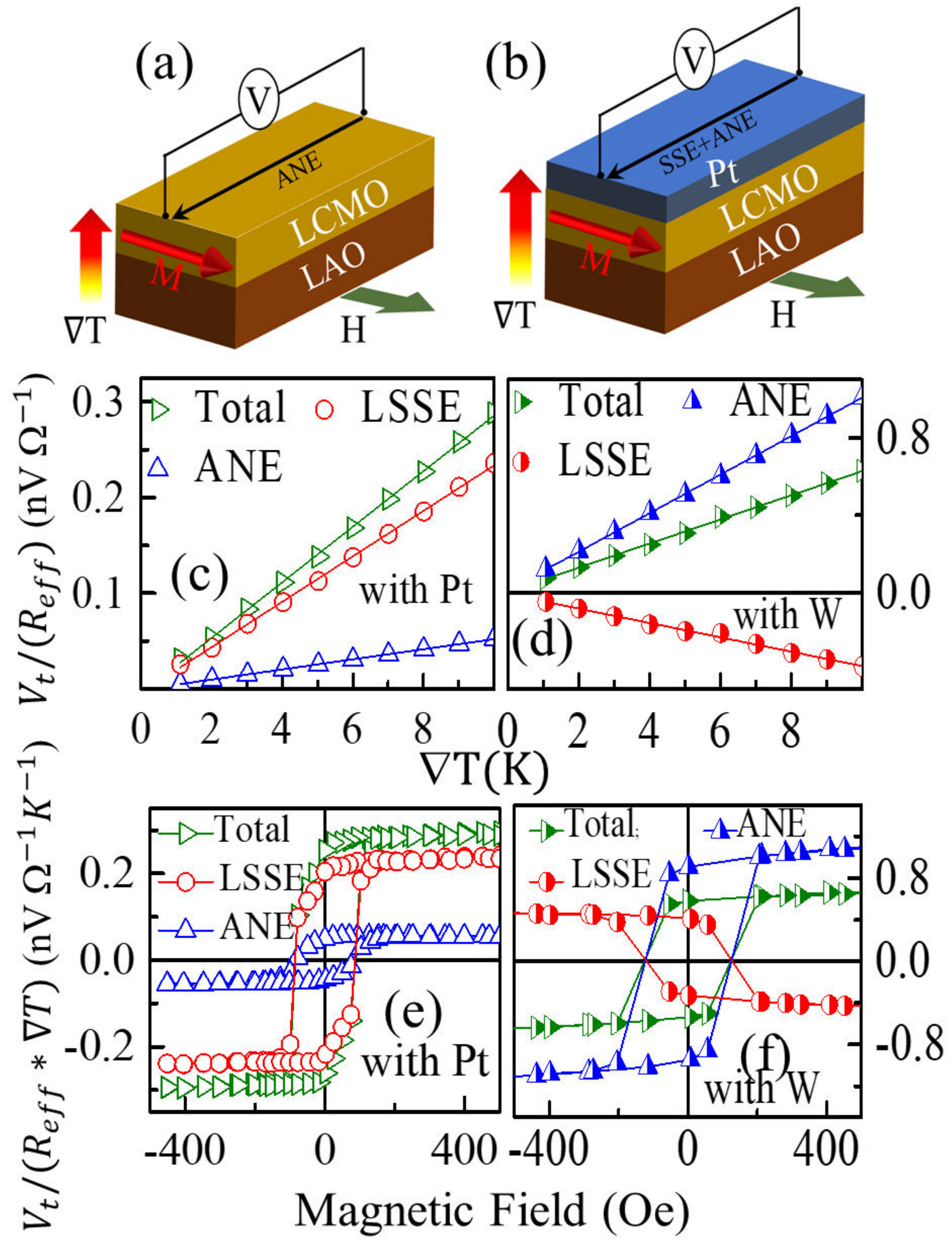}
	\caption{ (a) and (b) are schematic illustrations of the device for measuring the ANE and the total signal (LSSE+ANE) respectively. (c) and (d) depicts $\nabla T$ dependence of disentangled signals with a 10 nm Pt and a 10 nm W overlayer respectively. (e) and (f) depicts the $H$ dependence of these signals. The linearity as a function of $\nabla T$ and the reversal of the sign of the LSSE between the Pt and W layers confirms the intrinsic nature of the measured spin Seebeck signal. }
	\label{Fig1}
\end{figure} 

The investigations of material systems which are not as insulating as YIG are however plagued by contaminated SSE signals. These spurious signals arise as a consequence of the anomalous Nernst effect (ANE) in the magnetic sub-unit as well as the proximity induced ANE in the Pt layer, both of which - by virtue of the geometry of the LSSE - contribute additively to the measured voltage. The disentanglement of these additional contributions from the spin Seebeck signal is clearly imperative for an accurate description of this phenomena. This is especially so, since the agreement between theory and experiments remains tenuous, both in the vicinity of magnetic phase transitions, as well as in the low $T$ regime, where the magnonic spin current is expected to be relatively unaffected by phonons and other associated scattering mechanisms \cite{diniz2016microscopic,boona2016magnon}. 

Mixed valent manganites of the form La${_{1-x}}$$A{_x}$MnO${_3}$ (with $A$ being a divalent alkali metal),  where double exchange gives rise to ferromagnetism and concomitant metallicity and are of special interest owing to their high spin polarization and colossal magnetoresistance  \cite{salamon2001physics}. Interestingly, prior measurements of the LSSE on such optimally doped ferromagnetic  specimens have been contradictory to each other. For instance, a report using the transverse geometry failed to detect any voltage corresponding to the SSE in La${_{0.67}}$Sr${_{0.33}}$MnO${_3}$, and concluded that the observed magneto-thermoelectric voltages were entirely comprised of anomalous and planar Nernst effect signals \cite{bui2014anomalous}. On the other hand, a subsequent LSSE measurement on the same system suggested that more than 95\% of the thermally driven voltage arises from the magnonic spin current contribution alone, with the ANE contribution being barely discernible \cite{wu2017longitudinal}. In this letter, we report temperature dependent LSSE measurements of the closely related La${_{0.7}}$Ca${_{0.3}}$MnO${_3}$ system in an attempt to resolve this apparent contradiction. Interestingly, our results are at variance to both these reports, and also reveals a low-$T$ regime where the functional form of the observed SSE voltage matches well with an existing theoretical model.

Recently, a quantitative disentanglement of the SSE from both the possible spurious ANE contributions was reported \cite{bougiatioti2017quantitative}. Using a set of Pt-NiFe${_2}$O${_x}$/Ni${_{33}}$Fe${_{67}}$ bilayers (with sample resistances varying across 7 decades) and utilizing in-plane and out-of-plane measurement geometries, it was demonstrated that the proximity-induced ANE was a contributory factor only in the most metallic of specimens (with $\rho \approx10^{-7} \ohm m$). Since manganites are known to be \emph{dirty} metals (with resistivities of the order of $10^{-3}\ohm m$), a contamination of the LSSE signals by the proximity-induced ANE can be ruled out in our case. To disentangle the ANE and LSSE contributions, measurements were done in the standard LSSE geometry (Figure 1) on both - a bare LCMO film and a LCMO-Pt bilayer,  200 nm thick epitaxial thin films of the nominal composition La${_{0.7}}$Ca${_{0.3}}$MnO${_3}$ were grown on a LaAlO$_3$ $(100)$ substrate using pulsed laser deposition, and a 10 nm Pt (or a 13 nm W) film was coated on top of the magnetic layer using dc sputtering. The voltage measured across the the bare LCMO film comprises of the ANE alone ($V{_{ANE}}$), whereas that measured across the LCMO-Pt bilayer ($V{_{Total}}$) is a sum of the intrinsic magnonic contribution ($V{_{LSSE}}$) and the spurious contribution due to the ANE in the LCMO film ($V{_{ANE}}$).  Since these voltages are measured across different effective resistances, we report all the voltages in their normalized form ie, $V{_{Total}}$ / $V{_{LSSE}}$  and $V{_{ANE}}$ are normalized using the resistances of the LCMO-Pt(W) device and the bare LCMO film respectively.  Figure 1(c) and (d) depicts the linearity of the measured voltages (and the inferred  $V{_{LSSE}}$) as a function of $\nabla T$ with the cold end of the specimen fixed at 100K and the magnetic field at 500 Oe. Figure 1(c) and (d) depicts the hysteresis in these signals as a function of varying magnetic fields at a mean sample temperature of 105K. The magnitude and the sign of the measured $V{_{LSSE}}$ with the Pt and W spin-charge conversion layers are commensurate with the experimentally determined Spin Hall angles of these heavy metals \cite{wang2014scaling} - confirming the intrinsic nature of the inferred LSSE signals. 

\begin{figure}
	\centering
	\vspace{-0.1cm}
	\hspace{0cm}
	\includegraphics[scale=0.3]{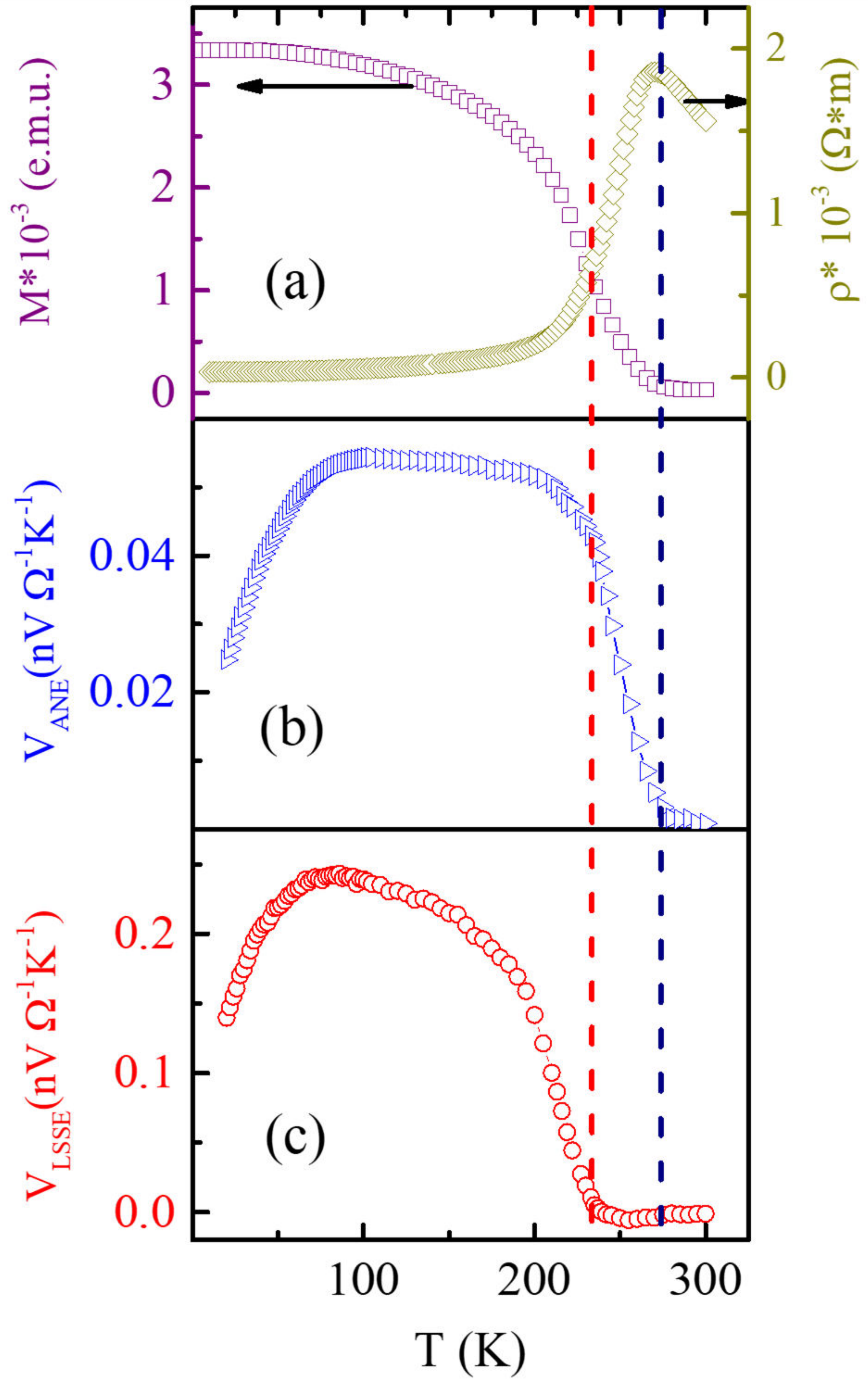}
	\caption{(a) depicts the temperature dependence of the magnetization and the film resistivity of the bare LCMO film. (b) and (c) depicts the temperature dependence of the $V_{ANE}$  and $V_{LSSE}$ respectively. The dashed lines indicate the difference between the onset of the two different signals. } 
	\label{Fig3}
\end{figure} 

The temperature dependence of the magnetization ($M$), the resistivity ($\rho$) and the individual ANE and LSSE contributions are depicted in Figure 2. Our LCMO specimen exhibits a double-exchange driven paramagnetic-ferromagnetic phase transition at 227 K, which is in line with the reported phase diagram \cite{huhtinen2002critical,salamon2001physics}. The onset of the transition in $V{_{ANE}}$ coincides exactly with the resistive transition, whereas the onset of $V{_{LSSE}}$ transition is observed deeper within the ordered phase. The temperature dependence of the ANE and SSE signals are also disparate. As a function of reducing temperature, the $V{_{ANE}}$ initially rises below the transition region - a consequence of both the increase in $M$ and reduction in $\rho$ - and below 100 K, $V{_{ANE}}$ falls as a function of temperature. By the Mott's relation, $V{_{ANE}}$ is expected to vary as $\frac{\pi ^2 k^2_BT}{3e} (\frac{\partial \sigma_{xy}}{\partial E})_{E_F} $, with $\sigma_{xy}$,  $k_B$,  $e$ and $E_F$ being the transverse electrical conductivity, the Boltzmann constant, the charge of an electron, and the energy at the Fermi level respectively \cite{miyasato2007crossover}. In the low-$T$ limit, as $M$ becomes invariant, $V{_{ANE}}$ is thus expected to vary linearly with $T$.  As is depicted in Figure 2(c) the temperature dependence of $V{_{LSSE}}$ is slightly different, with its increase  being steeper as a function of decreasing temperature. In similarity to that reported in other systems, a peak in $V{_{LSSE}}$($T$) is also observed at lower temperatures. This peak deep within the magnetically ordered regime is a defining feature of all $T$ dependent measurements of the LSSE and is suggested to arise as a consequence of the $T$ dependence of the magnon relaxation rates and population.
\begin{figure}
	\centering
	\vspace{-0.1cm}
	\hspace{-0.2cm}
	\includegraphics[scale=0.26]{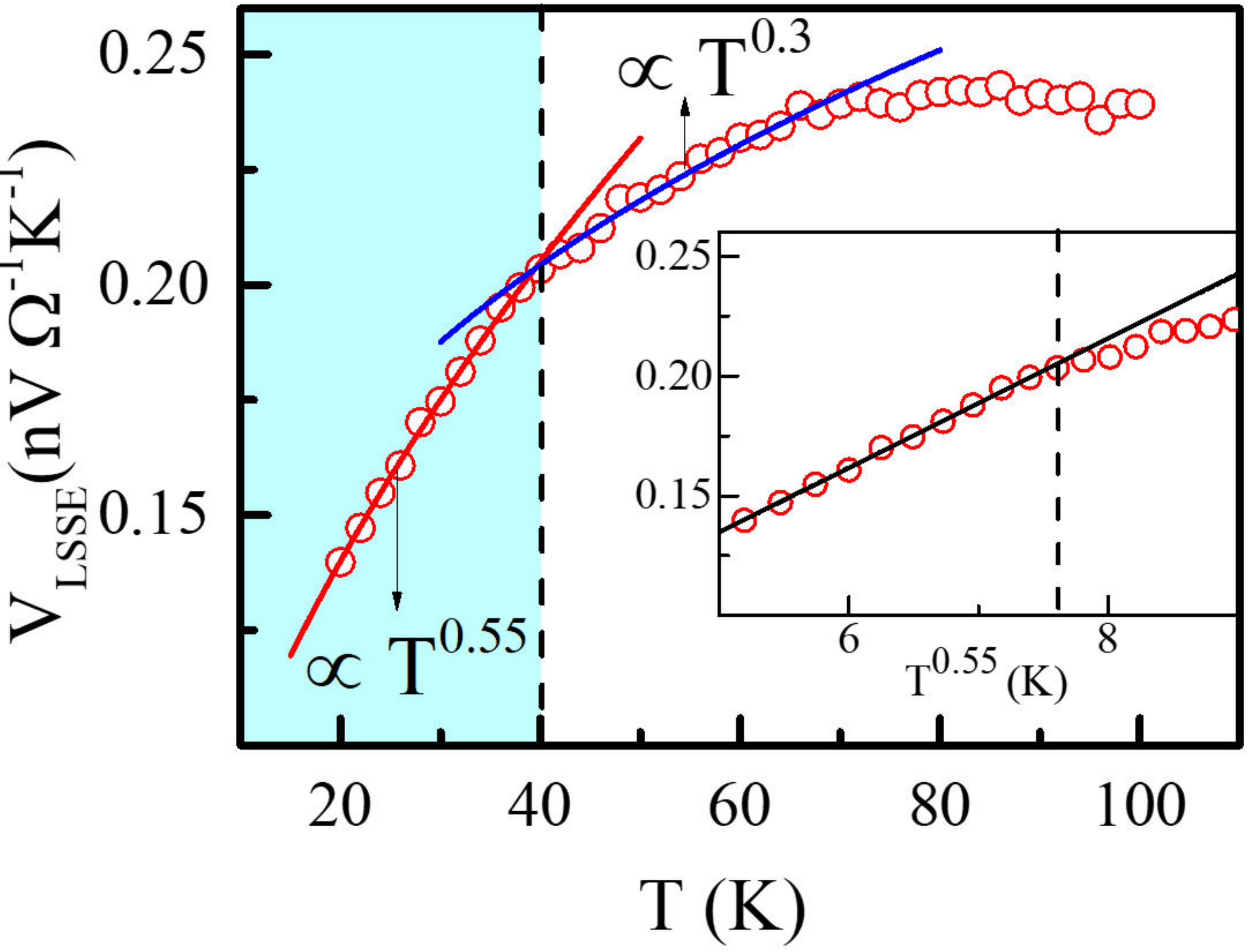}
	\caption{ Temperature dependence of  $V{_{LSSE}}$ is shown for temperatures below 100K. A power law fit with an exponent of $0.55$ is seen in the low temperature range $20K \leq T\leq 40K$ (red line), whereas at higher temperatures  ($40K \leq T\leq 70K$) an exponent of  ${0.3}$ is seen (blue line). The inset depicts a $T^{0.55}$ dependence of the  $V{_{LSSE}}$ at low temperatures. }
	\label{fig3}
\end{figure}  

Theoretical descriptions of the magnon-driven Spin Seebeck effect have relied on the mechanism of spin pumping caused by a finite difference between the effective magnon temperature in the ferromagnetic material and the electron (and phonon) temperature in the NM Pt layer. First proposed in the context of the transverse measurement geometry \cite{xiao2010theory}, this model was later modified for the LSSE geometry by Rezende and co-workers \cite{rezende2014magnon}. On an application of a thermal gradient $\vec{\nabla}T$, the spin current density ($\vec{J}_s$) flowing through the interface is given by $\vec{J}_s = D_s\vec{\nabla}T$. The spin Seebeck coefficient ($D_s$) has the form $D_s = \gamma\hbar k_B g{^{\uparrow\downarrow}}/(2\pi M_s V_a)$, where $\gamma$, $\hbar$, $g{^{\uparrow\downarrow}}$,$k_B$, $M_s$ and $V_a$ refer to the gyromagnetic ratio, the Planck constant, the spin-mixing conductance, the Boltzmann constant, the saturation magnetization and the magnetic coherence volume respectively \cite{xiao2010theory}. The temperature dependence of this spin current is thus primarily expected to arise as a consequence of the $T$ dependence of the magnetic coherence volume [$V_a = \frac{2}{3\zeta \frac{5}{2}} \frac{4\pi}{K_B}(\frac{D}{T})^{\frac{3}{2}}$, with $\zeta$ being the Riemann Zeta function and $D$ the spin stiffness constant], implying that $J_S \propto \frac{g{^{\uparrow\downarrow}}}{M_s}(\frac{T}{D})^{\frac{3}{2}} $. This spin current flows into the Pt layer, with a spin diffusion length $\lambda_{Pt}$ and generates a charge current $\vec{J_c} = \left(\frac{2e}{\hbar}\right) \theta_{Pt} \vec{J_s\times \sigma}$,where $e$, $\theta_{Pt}$ and $\sigma$ refer to the electronic charge, spin Hall angle of Pt and spin polarization of the conduction electrons in Pt respectively\cite{saitoh2006conversion}. The corresponding SSE voltage is given by $V_{LSSE} = R_d l_{Pt} \lambda_{Pt} \frac{2e}{\hbar} \theta_{Pt} \tanh \left( \frac{t_{Pt}}{2 \lambda_{Pt}}\right) J_s$, where $R_d, l_{Pt}$ and $t_{Pt} $ are the resistance between the contacts across Pt, length of the Pt bar and thickness of the Pt layer respectively \cite{arana2018spin,rezende2014magnon} .  Since  $\tanh \left( \frac{t_{Pt}}{2 \lambda_{Pt}} \right) \approx 1 $ in our case, and with $\lambda_{Pt} \propto T{^{-1}}$ \cite{marmion2014temperature} and $M_s$ being invariant at low temperatures, $V{_{LSSE}} \propto \frac{\theta{_{Pt}} g{^{\uparrow\downarrow}}T^{\frac{1}{2}}}{D^{\frac{3}{2}}}$ - implying that $V{_{LSSE}}$ should vary as $T{^{0.5}}$, at-least at low temperatures. However, prior measurements on the model YIG-Pt system have revealed a linear $T$ dependence, which has been attributed to the quadratic magnon dispersion in YIG \cite{jin2015effect}, and on the influence of the thermal conductivity in determining the functional form of $V{_{LSSE}}$ \cite{iguchi2017concomitant}.

Figure 3 depicts an expanded view of the $T$ dependence of  $V{_{LSSE}}$ as measured in the LCMO-Pt bilayer at low temperatures. We observe that at the lowest temperatures, $V{_{LSSE}}$ varies as $T^{0.55}$ - an observation which is in good agreement with the spin magnon theory. In excess of 40K, $V{_{LSSE}}$ exhibits a $T^{0.3}$ dependence,  presumably due to the additional phonon-magnon scattering processes. It has been suggested earlier that the low energy (or sub-thermal) magnons contribute disproportionately to the measured $V{_{LSSE}}$ signal, and that phonon mediated magnon scattering needs to be explicitly considered in describing the magnitude as well as the $T$ and $H$ dependence of the LSSE \cite{diniz2016microscopic, weiler2013experimental}. Earlier work on YIG-Pt devices have also suggested an intimate coupling between the thermal conductivity ($\kappa$) and the measured $V{_{LSSE}}$, reinforcing the importance of phonon-mediated processes in the LSSE \cite{iguchi2017concomitant}. We note that earlier thermal conductivity measurements on La${_{0.67}}$Ca${_{0.33}}$MnO${_3}$ have indicated that $\kappa$($T$) peaks at about 40K \cite{visser1997thermal}, and it is possible that the abrupt change in the measured $V{_{LSSE}}$ in this temperature regime stems from the change in $\kappa$($T$) of the magnetic layer.

Theory and experiments have also failed to reconcile with regard to the functional form of the SSE near the magnetic phase transition. The only prior measurements on YIG-Pt devices have reported that the SSE signal varied as $(T_c -T)^3$ \cite{uchida2014quantitative} or as $(T_c -T)^{1.5}$ \cite{wang2015spin} in the vicinity of the magnetic phase transition, whereas the magnetization scaled as $(T_c -T){^{0.5}}$, as is expected from mean field theory. However, subsequent theoretical investigations have predicted that the SSE should vary in consonance with the magnetization. For instance, an atomic numerical simulation considering the full spin wave spectrum of the ferrimagnetic YIG suggested that the SSE should have the same exponent as the magnetization \cite{barker2016thermal}. A time dependent Ginzburg-Landau analytical treatment of a (simple single-sublattice) ferromagnet also suggested that the SSE signal should vary along with the magnetization as $(T_c-T)^{0.5}$\cite{adachi2018spin}. Figure 4 depicts the temperature dependence of both $V{_{LSSE}}(T)$ as well as $M(T)$ in the vicinity of the para-ferromagnetic phase transition in LCMO. We obtain a critical exponent of $0.23$ for the magnetization which is in reasonable agreement with that reported earlier in bulk and thin film specimens of similar compositions \cite{pino2013single, huhtinen2002unconventional,kim2002tricritical}. Interestingly, we obtain an exponent of $0.7$ for $V{_{LSSE}}(T)$ in the same region. This is in remarkable similarity to that observed in YIG, where the LSSE exponent exceeds the magnetization one. It was speculated that the difference between the magnetization and LSSE exponents in the case of YIG/Pt could arise as a consequence of the \emph{ferri}magnetic nature of YIG, or other considerations like the magnetic surface anisotropy \cite{adachi2018spin}. Though spin wave spectra are expected to be material specific, the fact that we observe a similar discrepancy between the magnetization and LSSE exponents in the case of LCMO/Pt indicates this could be a generic feature of the LSSE signals in the vicinity of a magnetic phase transition.
\begin{figure}
	\centering
	\vspace{-0.1cm}
	\hspace{-0.1cm}
	\includegraphics[scale=0.31]{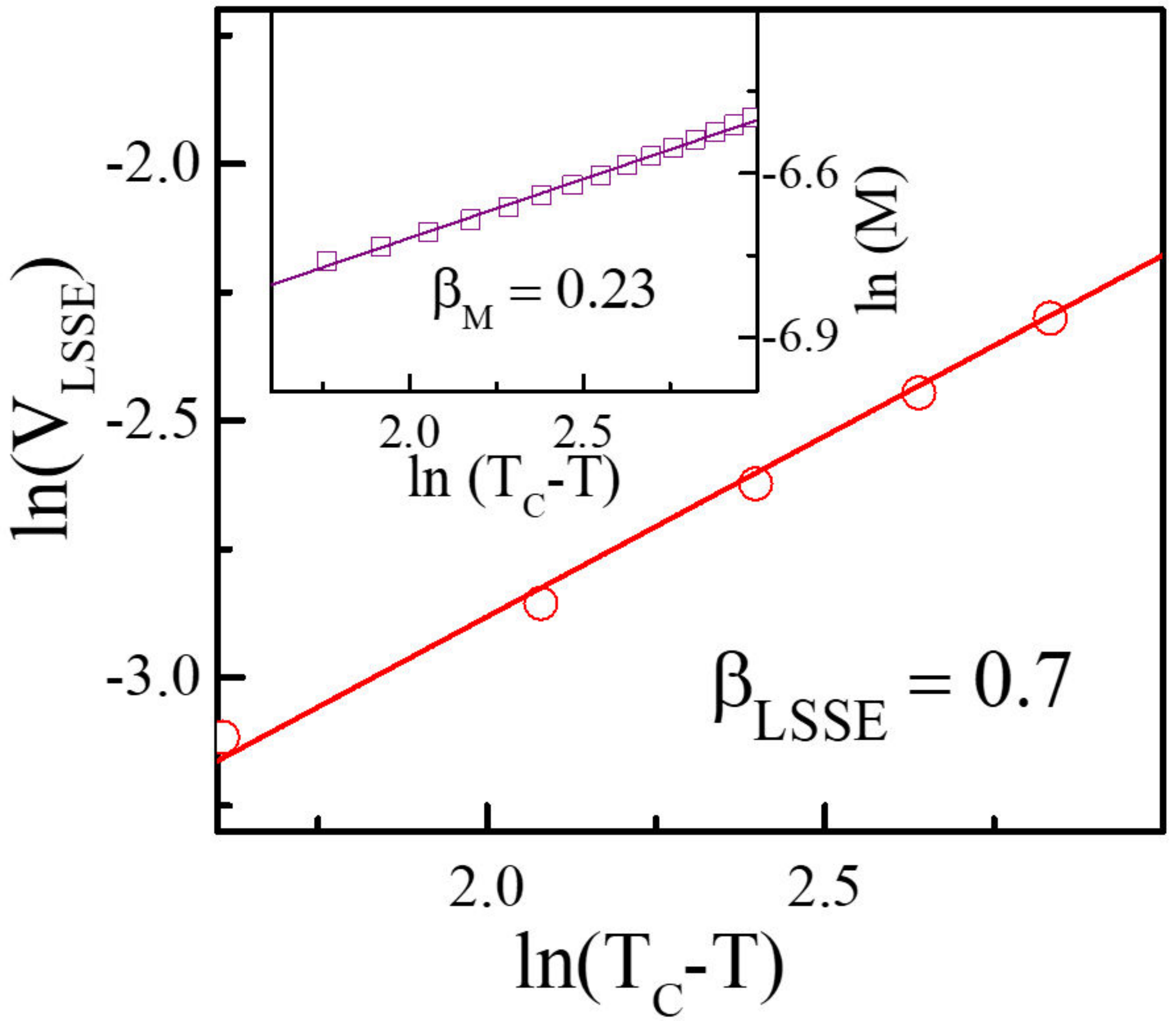}
	\caption{A double logarithmic plot of the $T_c-T$ dependence of the magnetization and the LSSE signal is depicted in the inset and the main panel respectively. The critical exponent of $V_{LSSE}$ is seen to be many times larger than that of the magnetization. }
	\label{Fig4}
\end{figure}  

It has been suggested that this difference in the $M$ and $V_{LSSE}$ critical exponents could be a consequence of the fact that the LSSE signal is not only a function of the static magnetic properties of the ferromagnet, but would also be determined by the $T$ dependence of other factors like the spin mixing conductance, the spin-diffusion length and the spin-Hall angle of the NM layer \cite{uchida2014quantitative}. We note that the intrinsic spin Hall angle of Pt is reported to be invariant in the temperature range of the LCMO transition, and hence is unlike to be a contributory factor \cite{isasa2015temperature}. The magnon-driven spin current density is expected to vary inversely with the magnetic coherence volume ($V_a$), which in turn has a $D^{3/2}$ dependence (with $D$ being the spin stiffness co-efficient). Interestingly, it has been reported that in LCMO this spin stiffness co-efficient has an anomalous $T$ dependence in the vicinity of the phase transition and does not fall to zero as $T \rightarrow T_c$ \cite{lynn1996unconventional}. This is presumably due to the finite magnetic correlations in the paramagnetic phase - a hallmark of the colossal magnetoresistive manganites.  It has also been demonstrated earlier, that the spin diffusion length of Pt is nearly invariant in the temperature range of the LCMO phase transition \cite{marmion2014temperature,isasa2015temperature}. With these factors being ruled out, it appears that the behaviour of $V_{LSSE}$($T$) in the vicinity of the magnetic phase transition could arise from the anomalous $T$ dependence of the magnetic coherence volume ($V_a$). An additional contribution could arise from the spin mixing conductance, whose $T$ dependence vis-a-vis the magnetization ($M$) remains to be investigated - at least in the case of LCMO. Clearly, more focused measurements across the para-ferromagnetic transitions in systems spanning across different universality classes would be required to gain further understanding of the $V_{LSSE}$($T$) in the vicinity of phase transitions. The temperature dependence of the spin mixing conductance could also probably explain why our measured $V_{LSSE}$($T$) matches well with the magnon-driven spin current model in the low-$T$ regime, whereas model systems like YIG-Pt do not exhibit this behavior. Recent ferromagnetic resonance (FMR) measurements on YIG have indicated a pronounced $T$ dependence of the FMR linewidth ($\Delta H$), implying a change in the Gilbert damping($\alpha$) at low temperatures since $\alpha = \frac{\gamma\Delta H}{2\omega}$, with $\gamma$ and $\omega$ being the gyromagnetic ratio and the microwave frequency \cite{jermain2017increased, maier2017temperature}. The spin mixing conductance being related to the Gilbert damping through the relation $g^{\uparrow\downarrow} = \frac{4\pi M_Sd_f}{g\mu_{\beta}} \Delta\alpha$ (with $M{_S}$, $d{_f}$ and $g$ being the  spontaneous magnetization, film thickness and the Lande's $g$ factor respectively), would thus be expected to have a considerable $T$ dependence  \cite{heinrich2011spin, mosendz2010quantifying, nakayama2012geometry}. On the other hand, early FMR measurements on La${_{0.7}}$Ca${_{0.3}}$MnO${_3}$ have indicated that $\Delta H$ term is relatively $T$ independent below 100K \cite{dyakonov2003ferromagnetic}, and we speculate that it is this relative $T$ invariance of the spin mixing conductance and other terms that allows us to observe the theoretically predicted $T^{0.5}$ dependence of $V_{LSSE}$($T$) in the low temperature regime. 

In summary, we report on temperature dependent measurements of the LSSE on an optimally doped La$_{0.7}$Ca$_{0.3}$MnO$_3$ system. After disentangling the ANE contribution we observe that these two thermal effects have disparate $T$ dependencies. In the low $T$ regime, $V_{LSSE}$($T$) varies as $T^{0.5}$, as is expected by the magnon-driven spin current model. In the para-ferromagnetic transition region, the $V_{LSSE}$ exhibits an exponent which is much larger than that of the magnetization - a feature which could be generic to most phase transitions of this nature. Our observations reinforces the need to individually ascertain the $T$ dependencies of contributory mechanisms which play a role in dictating both the magnitude of spin current as well as the extent of spin to charge conversion to understand the temperature dependent LSSE in strongly correlated materials.

AD acknowledges UGC, Govt. of India for providing financial support through a Senior Research Fellowship. AG acknowledges SERB, DST, Govt. of India for providing the financial support through a National Post-Doctoral Fellowship (PDF/2015/000599). The authors acknowledge funding support by the Department of Science and Technology (DST, Govt. of India) under the DST Nanomission Thematic Unit Program (SR/NM/TP-13/2016). 

\bibliography{Bibliography}

%merlin.mbs apsrev4-1.bst 2010-07-25 4.21a (PWD, AO, DPC) hacked
%Control: key (0)
%Control: author (8) initials jnrlst
%Control: editor formatted (1) identically to author
%Control: production of article title (-1) disabled
%Control: page (0) single
%Control: year (1) truncated
%Control: production of eprint (0) enabled
\begin{thebibliography}{45}%
\makeatletter
\providecommand \@ifxundefined [1]{%
 \@ifx{#1\undefined}
}%
\providecommand \@ifnum [1]{%
 \ifnum #1\expandafter \@firstoftwo
 \else \expandafter \@secondoftwo
 \fi
}%
\providecommand \@ifx [1]{%
 \ifx #1\expandafter \@firstoftwo
 \else \expandafter \@secondoftwo
 \fi
}%
\providecommand \natexlab [1]{#1}%
\providecommand \enquote  [1]{``#1''}%
\providecommand \bibnamefont  [1]{#1}%
\providecommand \bibfnamefont [1]{#1}%
\providecommand \citenamefont [1]{#1}%
\providecommand \href@noop [0]{\@secondoftwo}%
\providecommand \href [0]{\begingroup \@sanitize@url \@href}%
\providecommand \@href[1]{\@@startlink{#1}\@@href}%
\providecommand \@@href[1]{\endgroup#1\@@endlink}%
\providecommand \@sanitize@url [0]{\catcode `\\12\catcode `\$12\catcode
  `\&12\catcode `\#12\catcode `\^12\catcode `\_12\catcode `\%12\relax}%
\providecommand \@@startlink[1]{}%
\providecommand \@@endlink[0]{}%
\providecommand \url  [0]{\begingroup\@sanitize@url \@url }%
\providecommand \@url [1]{\endgroup\@href {#1}{\urlprefix }}%
\providecommand \urlprefix  [0]{URL }%
\providecommand \Eprint [0]{\href }%
\providecommand \doibase [0]{http://dx.doi.org/}%
\providecommand \selectlanguage [0]{\@gobble}%
\providecommand \bibinfo  [0]{\@secondoftwo}%
\providecommand \bibfield  [0]{\@secondoftwo}%
\providecommand \translation [1]{[#1]}%
\providecommand \BibitemOpen [0]{}%
\providecommand \bibitemStop [0]{}%
\providecommand \bibitemNoStop [0]{.\EOS\space}%
\providecommand \EOS [0]{\spacefactor3000\relax}%
\providecommand \BibitemShut  [1]{\csname bibitem#1\endcsname}%
\let\auto@bib@innerbib\@empty
%</preamble>
\bibitem [{\citenamefont {Uchida}\ \emph
  {et~al.}(2010{\natexlab{a}})\citenamefont {Uchida}, \citenamefont {Xiao},
  \citenamefont {Adachi}, \citenamefont {Ohe}, \citenamefont {Takahashi},
  \citenamefont {Ieda}, \citenamefont {Ota}, \citenamefont {Kajiwara},
  \citenamefont {Umezawa}, \citenamefont {Kawai} \emph
  {et~al.}}]{uchida2010spin}%
  \BibitemOpen
  \bibfield  {author} {\bibinfo {author} {\bibfnamefont {K.}~\bibnamefont
  {Uchida}}, \bibinfo {author} {\bibfnamefont {J.}~\bibnamefont {Xiao}},
  \bibinfo {author} {\bibfnamefont {H.}~\bibnamefont {Adachi}}, \bibinfo
  {author} {\bibfnamefont {J.-i.}\ \bibnamefont {Ohe}}, \bibinfo {author}
  {\bibfnamefont {S.}~\bibnamefont {Takahashi}}, \bibinfo {author}
  {\bibfnamefont {J.}~\bibnamefont {Ieda}}, \bibinfo {author} {\bibfnamefont
  {T.}~\bibnamefont {Ota}}, \bibinfo {author} {\bibfnamefont {Y.}~\bibnamefont
  {Kajiwara}}, \bibinfo {author} {\bibfnamefont {H.}~\bibnamefont {Umezawa}},
  \bibinfo {author} {\bibfnamefont {H.}~\bibnamefont {Kawai}},  \emph
  {et~al.},\ }\href@noop {} {\bibfield  {journal} {\bibinfo  {journal} {Nat.
  Mater.}\ }\textbf {\bibinfo {volume} {9}},\ \bibinfo {pages} {894} (\bibinfo
  {year} {2010}{\natexlab{a}})}\BibitemShut {NoStop}%
\bibitem [{\citenamefont {Uchida}\ \emph {et~al.}(2008)\citenamefont {Uchida},
  \citenamefont {Takahashi}, \citenamefont {Harii}, \citenamefont {Ieda},
  \citenamefont {Koshibae}, \citenamefont {Ando}, \citenamefont {Maekawa},\
  and\ \citenamefont {Saitoh}}]{uchida2008observation}%
  \BibitemOpen
  \bibfield  {author} {\bibinfo {author} {\bibfnamefont {K.}~\bibnamefont
  {Uchida}}, \bibinfo {author} {\bibfnamefont {S.}~\bibnamefont {Takahashi}},
  \bibinfo {author} {\bibfnamefont {K.}~\bibnamefont {Harii}}, \bibinfo
  {author} {\bibfnamefont {J.}~\bibnamefont {Ieda}}, \bibinfo {author}
  {\bibfnamefont {W.}~\bibnamefont {Koshibae}}, \bibinfo {author}
  {\bibfnamefont {K.}~\bibnamefont {Ando}}, \bibinfo {author} {\bibfnamefont
  {S.}~\bibnamefont {Maekawa}}, \ and\ \bibinfo {author} {\bibfnamefont
  {E.}~\bibnamefont {Saitoh}},\ }\href@noop {} {\bibfield  {journal} {\bibinfo
  {journal} {Nature}\ }\textbf {\bibinfo {volume} {455}},\ \bibinfo {pages}
  {778} (\bibinfo {year} {2008})}\BibitemShut {NoStop}%
\bibitem [{\citenamefont {Uchida}\ \emph
  {et~al.}(2010{\natexlab{b}})\citenamefont {Uchida}, \citenamefont {Adachi},
  \citenamefont {Ota}, \citenamefont {Nakayama}, \citenamefont {Maekawa},\ and\
  \citenamefont {Saitoh}}]{uchida2010observation}%
  \BibitemOpen
  \bibfield  {author} {\bibinfo {author} {\bibfnamefont {K.-i.}\ \bibnamefont
  {Uchida}}, \bibinfo {author} {\bibfnamefont {H.}~\bibnamefont {Adachi}},
  \bibinfo {author} {\bibfnamefont {T.}~\bibnamefont {Ota}}, \bibinfo {author}
  {\bibfnamefont {H.}~\bibnamefont {Nakayama}}, \bibinfo {author}
  {\bibfnamefont {S.}~\bibnamefont {Maekawa}}, \ and\ \bibinfo {author}
  {\bibfnamefont {E.}~\bibnamefont {Saitoh}},\ }\href@noop {} {\bibfield
  {journal} {\bibinfo  {journal} {Appl. Phys. Lett.}\ }\textbf {\bibinfo
  {volume} {97}},\ \bibinfo {pages} {172505} (\bibinfo {year}
  {2010}{\natexlab{b}})}\BibitemShut {NoStop}%
\bibitem [{\citenamefont {Kikkawa}\ \emph {et~al.}(2013)\citenamefont
  {Kikkawa}, \citenamefont {Uchida}, \citenamefont {Shiomi}, \citenamefont
  {Qiu}, \citenamefont {Hou}, \citenamefont {Tian}, \citenamefont {Nakayama},
  \citenamefont {Jin},\ and\ \citenamefont {Saitoh}}]{kikkawa2013longitudinal}%
  \BibitemOpen
  \bibfield  {author} {\bibinfo {author} {\bibfnamefont {T.}~\bibnamefont
  {Kikkawa}}, \bibinfo {author} {\bibfnamefont {K.}~\bibnamefont {Uchida}},
  \bibinfo {author} {\bibfnamefont {Y.}~\bibnamefont {Shiomi}}, \bibinfo
  {author} {\bibfnamefont {Z.}~\bibnamefont {Qiu}}, \bibinfo {author}
  {\bibfnamefont {D.}~\bibnamefont {Hou}}, \bibinfo {author} {\bibfnamefont
  {D.}~\bibnamefont {Tian}}, \bibinfo {author} {\bibfnamefont {H.}~\bibnamefont
  {Nakayama}}, \bibinfo {author} {\bibfnamefont {X.-F.}\ \bibnamefont {Jin}}, \
  and\ \bibinfo {author} {\bibfnamefont {E.}~\bibnamefont {Saitoh}},\
  }\href@noop {} {\bibfield  {journal} {\bibinfo  {journal} {Phys. Rev. Lett.}\
  }\textbf {\bibinfo {volume} {110}},\ \bibinfo {pages} {067207} (\bibinfo
  {year} {2013})}\BibitemShut {NoStop}%
\bibitem [{\citenamefont {Uchida}\ \emph {et~al.}(2014)\citenamefont {Uchida},
  \citenamefont {Kikkawa}, \citenamefont {Miura}, \citenamefont {Shiomi},\ and\
  \citenamefont {Saitoh}}]{uchida2014quantitative}%
  \BibitemOpen
  \bibfield  {author} {\bibinfo {author} {\bibfnamefont {K.-i.}\ \bibnamefont
  {Uchida}}, \bibinfo {author} {\bibfnamefont {T.}~\bibnamefont {Kikkawa}},
  \bibinfo {author} {\bibfnamefont {A.}~\bibnamefont {Miura}}, \bibinfo
  {author} {\bibfnamefont {J.}~\bibnamefont {Shiomi}}, \ and\ \bibinfo {author}
  {\bibfnamefont {E.}~\bibnamefont {Saitoh}},\ }\href@noop {} {\bibfield
  {journal} {\bibinfo  {journal} {Phys. Rev. X}\ }\textbf {\bibinfo {volume}
  {4}},\ \bibinfo {pages} {041023} (\bibinfo {year} {2014})}\BibitemShut
  {NoStop}%
\bibitem [{\citenamefont {Giles}\ \emph {et~al.}(2017)\citenamefont {Giles},
  \citenamefont {Yang}, \citenamefont {Jamison}, \citenamefont {Gomez-Perez},
  \citenamefont {V{\'e}lez}, \citenamefont {Hueso}, \citenamefont {Casanova},\
  and\ \citenamefont {Myers}}]{giles2017thermally}%
  \BibitemOpen
  \bibfield  {author} {\bibinfo {author} {\bibfnamefont {B.~L.}\ \bibnamefont
  {Giles}}, \bibinfo {author} {\bibfnamefont {Z.}~\bibnamefont {Yang}},
  \bibinfo {author} {\bibfnamefont {J.~S.}\ \bibnamefont {Jamison}}, \bibinfo
  {author} {\bibfnamefont {J.~M.}\ \bibnamefont {Gomez-Perez}}, \bibinfo
  {author} {\bibfnamefont {S.}~\bibnamefont {V{\'e}lez}}, \bibinfo {author}
  {\bibfnamefont {L.~E.}\ \bibnamefont {Hueso}}, \bibinfo {author}
  {\bibfnamefont {F.}~\bibnamefont {Casanova}}, \ and\ \bibinfo {author}
  {\bibfnamefont {R.~C.}\ \bibnamefont {Myers}},\ }\href@noop {} {\bibfield
  {journal} {\bibinfo  {journal} {Phys. Rev. B}\ }\textbf {\bibinfo {volume}
  {96}},\ \bibinfo {pages} {180412} (\bibinfo {year} {2017})}\BibitemShut
  {NoStop}%
\bibitem [{\citenamefont {Meier}\ \emph {et~al.}(2015)\citenamefont {Meier},
  \citenamefont {Reinhardt}, \citenamefont {Van~Straaten}, \citenamefont
  {Klewe}, \citenamefont {Althammer}, \citenamefont {Schreier}, \citenamefont
  {Goennenwein}, \citenamefont {Gupta}, \citenamefont {Schmid}, \citenamefont
  {Back} \emph {et~al.}}]{meier2015longitudinal}%
  \BibitemOpen
  \bibfield  {author} {\bibinfo {author} {\bibfnamefont {D.}~\bibnamefont
  {Meier}}, \bibinfo {author} {\bibfnamefont {D.}~\bibnamefont {Reinhardt}},
  \bibinfo {author} {\bibfnamefont {M.}~\bibnamefont {Van~Straaten}}, \bibinfo
  {author} {\bibfnamefont {C.}~\bibnamefont {Klewe}}, \bibinfo {author}
  {\bibfnamefont {M.}~\bibnamefont {Althammer}}, \bibinfo {author}
  {\bibfnamefont {M.}~\bibnamefont {Schreier}}, \bibinfo {author}
  {\bibfnamefont {S.~T.}\ \bibnamefont {Goennenwein}}, \bibinfo {author}
  {\bibfnamefont {A.}~\bibnamefont {Gupta}}, \bibinfo {author} {\bibfnamefont
  {M.}~\bibnamefont {Schmid}}, \bibinfo {author} {\bibfnamefont {C.~H.}\
  \bibnamefont {Back}},  \emph {et~al.},\ }\href@noop {} {\bibfield  {journal}
  {\bibinfo  {journal} {Nat. Commun.}\ }\textbf {\bibinfo {volume} {6}},\
  \bibinfo {pages} {8211} (\bibinfo {year} {2015})}\BibitemShut {NoStop}%
\bibitem [{\citenamefont {Saitoh}\ \emph {et~al.}(2006)\citenamefont {Saitoh},
  \citenamefont {Ueda}, \citenamefont {Miyajima},\ and\ \citenamefont
  {Tatara}}]{saitoh2006conversion}%
  \BibitemOpen
  \bibfield  {author} {\bibinfo {author} {\bibfnamefont {E.}~\bibnamefont
  {Saitoh}}, \bibinfo {author} {\bibfnamefont {M.}~\bibnamefont {Ueda}},
  \bibinfo {author} {\bibfnamefont {H.}~\bibnamefont {Miyajima}}, \ and\
  \bibinfo {author} {\bibfnamefont {G.}~\bibnamefont {Tatara}},\ }\href@noop {}
  {\bibfield  {journal} {\bibinfo  {journal} {Appl. Phys. Lett.}\ }\textbf
  {\bibinfo {volume} {88}},\ \bibinfo {pages} {182509} (\bibinfo {year}
  {2006})}\BibitemShut {NoStop}%
\bibitem [{\citenamefont {Iguchi}\ \emph {et~al.}(2017)\citenamefont {Iguchi},
  \citenamefont {Uchida}, \citenamefont {Daimon},\ and\ \citenamefont
  {Saitoh}}]{iguchi2017concomitant}%
  \BibitemOpen
  \bibfield  {author} {\bibinfo {author} {\bibfnamefont {R.}~\bibnamefont
  {Iguchi}}, \bibinfo {author} {\bibfnamefont {K.-i.}\ \bibnamefont {Uchida}},
  \bibinfo {author} {\bibfnamefont {S.}~\bibnamefont {Daimon}}, \ and\ \bibinfo
  {author} {\bibfnamefont {E.}~\bibnamefont {Saitoh}},\ }\href@noop {}
  {\bibfield  {journal} {\bibinfo  {journal} {Phys. Rev. B}\ }\textbf {\bibinfo
  {volume} {95}},\ \bibinfo {pages} {174401} (\bibinfo {year}
  {2017})}\BibitemShut {NoStop}%
\bibitem [{\citenamefont {Jin}\ \emph {et~al.}(2015)\citenamefont {Jin},
  \citenamefont {Boona}, \citenamefont {Yang}, \citenamefont {Myers},\ and\
  \citenamefont {Heremans}}]{jin2015effect}%
  \BibitemOpen
  \bibfield  {author} {\bibinfo {author} {\bibfnamefont {H.}~\bibnamefont
  {Jin}}, \bibinfo {author} {\bibfnamefont {S.~R.}\ \bibnamefont {Boona}},
  \bibinfo {author} {\bibfnamefont {Z.}~\bibnamefont {Yang}}, \bibinfo {author}
  {\bibfnamefont {R.~C.}\ \bibnamefont {Myers}}, \ and\ \bibinfo {author}
  {\bibfnamefont {J.~P.}\ \bibnamefont {Heremans}},\ }\href@noop {} {\bibfield
  {journal} {\bibinfo  {journal} {Phys. Rev. B}\ }\textbf {\bibinfo {volume}
  {92}},\ \bibinfo {pages} {054436} (\bibinfo {year} {2015})}\BibitemShut
  {NoStop}%
\bibitem [{\citenamefont {Wang}\ \emph {et~al.}(2015)\citenamefont {Wang},
  \citenamefont {Zou}, \citenamefont {Zhang}, \citenamefont {Cai},
  \citenamefont {Wang}, \citenamefont {Shen},\ and\ \citenamefont
  {Sun}}]{wang2015spin}%
  \BibitemOpen
  \bibfield  {author} {\bibinfo {author} {\bibfnamefont {S.}~\bibnamefont
  {Wang}}, \bibinfo {author} {\bibfnamefont {L.}~\bibnamefont {Zou}}, \bibinfo
  {author} {\bibfnamefont {X.}~\bibnamefont {Zhang}}, \bibinfo {author}
  {\bibfnamefont {J.}~\bibnamefont {Cai}}, \bibinfo {author} {\bibfnamefont
  {S.}~\bibnamefont {Wang}}, \bibinfo {author} {\bibfnamefont {B.}~\bibnamefont
  {Shen}}, \ and\ \bibinfo {author} {\bibfnamefont {J.}~\bibnamefont {Sun}},\
  }\href@noop {} {\bibfield  {journal} {\bibinfo  {journal} {Nanoscale}\
  }\textbf {\bibinfo {volume} {7}},\ \bibinfo {pages} {17812} (\bibinfo {year}
  {2015})}\BibitemShut {NoStop}%
\bibitem [{\citenamefont {Kikkawa}\ \emph {et~al.}(2015)\citenamefont
  {Kikkawa}, \citenamefont {Uchida}, \citenamefont {Daimon}, \citenamefont
  {Qiu}, \citenamefont {Shiomi},\ and\ \citenamefont
  {Saitoh}}]{kikkawa2015critical}%
  \BibitemOpen
  \bibfield  {author} {\bibinfo {author} {\bibfnamefont {T.}~\bibnamefont
  {Kikkawa}}, \bibinfo {author} {\bibfnamefont {K.-i.}\ \bibnamefont {Uchida}},
  \bibinfo {author} {\bibfnamefont {S.}~\bibnamefont {Daimon}}, \bibinfo
  {author} {\bibfnamefont {Z.}~\bibnamefont {Qiu}}, \bibinfo {author}
  {\bibfnamefont {Y.}~\bibnamefont {Shiomi}}, \ and\ \bibinfo {author}
  {\bibfnamefont {E.}~\bibnamefont {Saitoh}},\ }\href@noop {} {\bibfield
  {journal} {\bibinfo  {journal} {Phys. Rev. B}\ }\textbf {\bibinfo {volume}
  {92}},\ \bibinfo {pages} {064413} (\bibinfo {year} {2015})}\BibitemShut
  {NoStop}%
\bibitem [{\citenamefont {Ritzmann}\ \emph {et~al.}(2015)\citenamefont
  {Ritzmann}, \citenamefont {Hinzke}, \citenamefont {Kehlberger}, \citenamefont
  {Guo}, \citenamefont {Kl{\"a}ui},\ and\ \citenamefont
  {Nowak}}]{ritzmann2015magnetic}%
  \BibitemOpen
  \bibfield  {author} {\bibinfo {author} {\bibfnamefont {U.}~\bibnamefont
  {Ritzmann}}, \bibinfo {author} {\bibfnamefont {D.}~\bibnamefont {Hinzke}},
  \bibinfo {author} {\bibfnamefont {A.}~\bibnamefont {Kehlberger}}, \bibinfo
  {author} {\bibfnamefont {E.-J.}\ \bibnamefont {Guo}}, \bibinfo {author}
  {\bibfnamefont {M.}~\bibnamefont {Kl{\"a}ui}}, \ and\ \bibinfo {author}
  {\bibfnamefont {U.}~\bibnamefont {Nowak}},\ }\href@noop {} {\bibfield
  {journal} {\bibinfo  {journal} {Phys. Rev. B}\ }\textbf {\bibinfo {volume}
  {92}},\ \bibinfo {pages} {174411} (\bibinfo {year} {2015})}\BibitemShut
  {NoStop}%
\bibitem [{\citenamefont {Kehlberger}\ \emph {et~al.}(2015)\citenamefont
  {Kehlberger}, \citenamefont {Ritzmann}, \citenamefont {Hinzke}, \citenamefont
  {Guo}, \citenamefont {Cramer}, \citenamefont {Jakob}, \citenamefont
  {Onbasli}, \citenamefont {Kim}, \citenamefont {Ross}, \citenamefont
  {Jungfleisch} \emph {et~al.}}]{kehlberger2015length}%
  \BibitemOpen
  \bibfield  {author} {\bibinfo {author} {\bibfnamefont {A.}~\bibnamefont
  {Kehlberger}}, \bibinfo {author} {\bibfnamefont {U.}~\bibnamefont
  {Ritzmann}}, \bibinfo {author} {\bibfnamefont {D.}~\bibnamefont {Hinzke}},
  \bibinfo {author} {\bibfnamefont {E.-J.}\ \bibnamefont {Guo}}, \bibinfo
  {author} {\bibfnamefont {J.}~\bibnamefont {Cramer}}, \bibinfo {author}
  {\bibfnamefont {G.}~\bibnamefont {Jakob}}, \bibinfo {author} {\bibfnamefont
  {M.~C.}\ \bibnamefont {Onbasli}}, \bibinfo {author} {\bibfnamefont {D.~H.}\
  \bibnamefont {Kim}}, \bibinfo {author} {\bibfnamefont {C.~A.}\ \bibnamefont
  {Ross}}, \bibinfo {author} {\bibfnamefont {M.~B.}\ \bibnamefont
  {Jungfleisch}},  \emph {et~al.},\ }\href@noop {} {\bibfield  {journal}
  {\bibinfo  {journal} {Phys. Rev. Lett.}\ }\textbf {\bibinfo {volume} {115}},\
  \bibinfo {pages} {096602} (\bibinfo {year} {2015})}\BibitemShut {NoStop}%
\bibitem [{\citenamefont {Guo}\ \emph {et~al.}(2016)\citenamefont {Guo},
  \citenamefont {Cramer}, \citenamefont {Kehlberger}, \citenamefont {Ferguson},
  \citenamefont {MacLaren}, \citenamefont {Jakob},\ and\ \citenamefont
  {Kl{\"a}ui}}]{guo2016influence}%
  \BibitemOpen
  \bibfield  {author} {\bibinfo {author} {\bibfnamefont {E.-J.}\ \bibnamefont
  {Guo}}, \bibinfo {author} {\bibfnamefont {J.}~\bibnamefont {Cramer}},
  \bibinfo {author} {\bibfnamefont {A.}~\bibnamefont {Kehlberger}}, \bibinfo
  {author} {\bibfnamefont {C.~A.}\ \bibnamefont {Ferguson}}, \bibinfo {author}
  {\bibfnamefont {D.~A.}\ \bibnamefont {MacLaren}}, \bibinfo {author}
  {\bibfnamefont {G.}~\bibnamefont {Jakob}}, \ and\ \bibinfo {author}
  {\bibfnamefont {M.}~\bibnamefont {Kl{\"a}ui}},\ }\href@noop {} {\bibfield
  {journal} {\bibinfo  {journal} {Phys. Rev. X}\ }\textbf {\bibinfo {volume}
  {6}},\ \bibinfo {pages} {031012} (\bibinfo {year} {2016})}\BibitemShut
  {NoStop}%
\bibitem [{\citenamefont {Ramos}\ \emph {et~al.}(2013)\citenamefont {Ramos},
  \citenamefont {Kikkawa}, \citenamefont {Uchida}, \citenamefont {Adachi},
  \citenamefont {Lucas}, \citenamefont {Aguirre}, \citenamefont {Algarabel},
  \citenamefont {Morell{\'o}n}, \citenamefont {Maekawa}, \citenamefont {Saitoh}
  \emph {et~al.}}]{ramos2013observation}%
  \BibitemOpen
  \bibfield  {author} {\bibinfo {author} {\bibfnamefont {R.}~\bibnamefont
  {Ramos}}, \bibinfo {author} {\bibfnamefont {T.}~\bibnamefont {Kikkawa}},
  \bibinfo {author} {\bibfnamefont {K.}~\bibnamefont {Uchida}}, \bibinfo
  {author} {\bibfnamefont {H.}~\bibnamefont {Adachi}}, \bibinfo {author}
  {\bibfnamefont {I.}~\bibnamefont {Lucas}}, \bibinfo {author} {\bibfnamefont
  {M.}~\bibnamefont {Aguirre}}, \bibinfo {author} {\bibfnamefont
  {P.}~\bibnamefont {Algarabel}}, \bibinfo {author} {\bibfnamefont
  {L.}~\bibnamefont {Morell{\'o}n}}, \bibinfo {author} {\bibfnamefont
  {S.}~\bibnamefont {Maekawa}}, \bibinfo {author} {\bibfnamefont
  {E.}~\bibnamefont {Saitoh}},  \emph {et~al.},\ }\href@noop {} {\bibfield
  {journal} {\bibinfo  {journal} {Appl. Phys. Lett.}\ }\textbf {\bibinfo
  {volume} {102}},\ \bibinfo {pages} {072413} (\bibinfo {year}
  {2013})}\BibitemShut {NoStop}%
\bibitem [{\citenamefont {Gepr{\"a}gs}\ \emph {et~al.}(2016)\citenamefont
  {Gepr{\"a}gs}, \citenamefont {Kehlberger}, \citenamefont {Della~Coletta},
  \citenamefont {Qiu}, \citenamefont {Guo}, \citenamefont {Schulz},
  \citenamefont {Mix}, \citenamefont {Meyer}, \citenamefont {Kamra},
  \citenamefont {Althammer} \emph {et~al.}}]{geprags2016origin}%
  \BibitemOpen
  \bibfield  {author} {\bibinfo {author} {\bibfnamefont {S.}~\bibnamefont
  {Gepr{\"a}gs}}, \bibinfo {author} {\bibfnamefont {A.}~\bibnamefont
  {Kehlberger}}, \bibinfo {author} {\bibfnamefont {F.}~\bibnamefont
  {Della~Coletta}}, \bibinfo {author} {\bibfnamefont {Z.}~\bibnamefont {Qiu}},
  \bibinfo {author} {\bibfnamefont {E.-J.}\ \bibnamefont {Guo}}, \bibinfo
  {author} {\bibfnamefont {T.}~\bibnamefont {Schulz}}, \bibinfo {author}
  {\bibfnamefont {C.}~\bibnamefont {Mix}}, \bibinfo {author} {\bibfnamefont
  {S.}~\bibnamefont {Meyer}}, \bibinfo {author} {\bibfnamefont
  {A.}~\bibnamefont {Kamra}}, \bibinfo {author} {\bibfnamefont
  {M.}~\bibnamefont {Althammer}},  \emph {et~al.},\ }\href@noop {} {\bibfield
  {journal} {\bibinfo  {journal} {Nat. Commun.}\ }\textbf {\bibinfo {volume}
  {7}},\ \bibinfo {pages} {10452} (\bibinfo {year} {2016})}\BibitemShut
  {NoStop}%
\bibitem [{\citenamefont {Wu}\ \emph {et~al.}(2017)\citenamefont {Wu},
  \citenamefont {Luo}, \citenamefont {Lin},\ and\ \citenamefont
  {Huang}}]{wu2017longitudinal}%
  \BibitemOpen
  \bibfield  {author} {\bibinfo {author} {\bibfnamefont {B.}~\bibnamefont
  {Wu}}, \bibinfo {author} {\bibfnamefont {G.}~\bibnamefont {Luo}}, \bibinfo
  {author} {\bibfnamefont {J.}~\bibnamefont {Lin}}, \ and\ \bibinfo {author}
  {\bibfnamefont {S.}~\bibnamefont {Huang}},\ }\href@noop {} {\bibfield
  {journal} {\bibinfo  {journal} {Phys. Rev. B}\ }\textbf {\bibinfo {volume}
  {96}},\ \bibinfo {pages} {060402} (\bibinfo {year} {2017})}\BibitemShut
  {NoStop}%
\bibitem [{\citenamefont {Diniz}\ and\ \citenamefont
  {Costa}(2016)}]{diniz2016microscopic}%
  \BibitemOpen
  \bibfield  {author} {\bibinfo {author} {\bibfnamefont {I.}~\bibnamefont
  {Diniz}}\ and\ \bibinfo {author} {\bibfnamefont {A.}~\bibnamefont {Costa}},\
  }\href@noop {} {\bibfield  {journal} {\bibinfo  {journal} {New J. Phys.}\
  }\textbf {\bibinfo {volume} {18}},\ \bibinfo {pages} {052002} (\bibinfo
  {year} {2016})}\BibitemShut {NoStop}%
\bibitem [{\citenamefont {Boona}(2016)}]{boona2016magnon}%
  \BibitemOpen
  \bibfield  {author} {\bibinfo {author} {\bibfnamefont {S.~R.}\ \bibnamefont
  {Boona}},\ }\href@noop {} {\bibfield  {journal} {\bibinfo  {journal} {New J.
  Phys.}\ }\textbf {\bibinfo {volume} {18}},\ \bibinfo {pages} {061002}
  (\bibinfo {year} {2016})}\BibitemShut {NoStop}%
\bibitem [{\citenamefont {Salamon}\ and\ \citenamefont
  {Jaime}(2001)}]{salamon2001physics}%
  \BibitemOpen
  \bibfield  {author} {\bibinfo {author} {\bibfnamefont {M.~B.}\ \bibnamefont
  {Salamon}}\ and\ \bibinfo {author} {\bibfnamefont {M.}~\bibnamefont
  {Jaime}},\ }\href@noop {} {\bibfield  {journal} {\bibinfo  {journal} {Rev.
  Mod. Phys.}\ }\textbf {\bibinfo {volume} {73}},\ \bibinfo {pages} {583}
  (\bibinfo {year} {2001})}\BibitemShut {NoStop}%
\bibitem [{\citenamefont {Bui}\ and\ \citenamefont
  {Rivadulla}(2014)}]{bui2014anomalous}%
  \BibitemOpen
  \bibfield  {author} {\bibinfo {author} {\bibfnamefont {C.~T.}\ \bibnamefont
  {Bui}}\ and\ \bibinfo {author} {\bibfnamefont {F.}~\bibnamefont
  {Rivadulla}},\ }\href@noop {} {\bibfield  {journal} {\bibinfo  {journal}
  {Phys. Rev. B}\ }\textbf {\bibinfo {volume} {90}},\ \bibinfo {pages} {100403}
  (\bibinfo {year} {2014})}\BibitemShut {NoStop}%
\bibitem [{\citenamefont {Bougiatioti}\ \emph {et~al.}(2017)\citenamefont
  {Bougiatioti}, \citenamefont {Klewe}, \citenamefont {Meier}, \citenamefont
  {Manos}, \citenamefont {Kuschel}, \citenamefont {Wollschl{\"a}ger},
  \citenamefont {Bouchenoire}, \citenamefont {Brown}, \citenamefont
  {Schmalhorst}, \citenamefont {Reiss} \emph
  {et~al.}}]{bougiatioti2017quantitative}%
  \BibitemOpen
  \bibfield  {author} {\bibinfo {author} {\bibfnamefont {P.}~\bibnamefont
  {Bougiatioti}}, \bibinfo {author} {\bibfnamefont {C.}~\bibnamefont {Klewe}},
  \bibinfo {author} {\bibfnamefont {D.}~\bibnamefont {Meier}}, \bibinfo
  {author} {\bibfnamefont {O.}~\bibnamefont {Manos}}, \bibinfo {author}
  {\bibfnamefont {O.}~\bibnamefont {Kuschel}}, \bibinfo {author} {\bibfnamefont
  {J.}~\bibnamefont {Wollschl{\"a}ger}}, \bibinfo {author} {\bibfnamefont
  {L.}~\bibnamefont {Bouchenoire}}, \bibinfo {author} {\bibfnamefont {S.~D.}\
  \bibnamefont {Brown}}, \bibinfo {author} {\bibfnamefont {J.-M.}\ \bibnamefont
  {Schmalhorst}}, \bibinfo {author} {\bibfnamefont {G.}~\bibnamefont {Reiss}},
  \emph {et~al.},\ }\href@noop {} {\bibfield  {journal} {\bibinfo  {journal}
  {Phys. Rev. Lett.}\ }\textbf {\bibinfo {volume} {119}},\ \bibinfo {pages}
  {227205} (\bibinfo {year} {2017})}\BibitemShut {NoStop}%
\bibitem [{\citenamefont {Wang}\ \emph {et~al.}(2014)\citenamefont {Wang},
  \citenamefont {Du}, \citenamefont {Pu}, \citenamefont {Adur}, \citenamefont
  {Hammel},\ and\ \citenamefont {Yang}}]{wang2014scaling}%
  \BibitemOpen
  \bibfield  {author} {\bibinfo {author} {\bibfnamefont {H.}~\bibnamefont
  {Wang}}, \bibinfo {author} {\bibfnamefont {C.}~\bibnamefont {Du}}, \bibinfo
  {author} {\bibfnamefont {Y.}~\bibnamefont {Pu}}, \bibinfo {author}
  {\bibfnamefont {R.}~\bibnamefont {Adur}}, \bibinfo {author} {\bibfnamefont
  {P.~C.}\ \bibnamefont {Hammel}}, \ and\ \bibinfo {author} {\bibfnamefont
  {F.}~\bibnamefont {Yang}},\ }\href@noop {} {\bibfield  {journal} {\bibinfo
  {journal} {Phys. Rev. Lett.}\ }\textbf {\bibinfo {volume} {112}},\ \bibinfo
  {pages} {197201} (\bibinfo {year} {2014})}\BibitemShut {NoStop}%
\bibitem [{\citenamefont {Huhtinen}\ \emph
  {et~al.}(2002{\natexlab{a}})\citenamefont {Huhtinen}, \citenamefont {Laiho},
  \citenamefont {Lisunov}, \citenamefont {Stamov},\ and\ \citenamefont
  {Zakhvalinskii}}]{huhtinen2002critical}%
  \BibitemOpen
  \bibfield  {author} {\bibinfo {author} {\bibfnamefont {H.}~\bibnamefont
  {Huhtinen}}, \bibinfo {author} {\bibfnamefont {R.}~\bibnamefont {Laiho}},
  \bibinfo {author} {\bibfnamefont {K.}~\bibnamefont {Lisunov}}, \bibinfo
  {author} {\bibfnamefont {V.}~\bibnamefont {Stamov}}, \ and\ \bibinfo {author}
  {\bibfnamefont {V.}~\bibnamefont {Zakhvalinskii}},\ }\href@noop {} {\bibfield
   {journal} {\bibinfo  {journal} {J. Magn. Magn. Mater.}\ }\textbf {\bibinfo
  {volume} {238}},\ \bibinfo {pages} {160} (\bibinfo {year}
  {2002}{\natexlab{a}})}\BibitemShut {NoStop}%
\bibitem [{\citenamefont {Miyasato}\ \emph {et~al.}(2007)\citenamefont
  {Miyasato}, \citenamefont {Abe}, \citenamefont {Fujii}, \citenamefont
  {Asamitsu}, \citenamefont {Onoda}, \citenamefont {Onose}, \citenamefont
  {Nagaosa},\ and\ \citenamefont {Tokura}}]{miyasato2007crossover}%
  \BibitemOpen
  \bibfield  {author} {\bibinfo {author} {\bibfnamefont {T.}~\bibnamefont
  {Miyasato}}, \bibinfo {author} {\bibfnamefont {N.}~\bibnamefont {Abe}},
  \bibinfo {author} {\bibfnamefont {T.}~\bibnamefont {Fujii}}, \bibinfo
  {author} {\bibfnamefont {A.}~\bibnamefont {Asamitsu}}, \bibinfo {author}
  {\bibfnamefont {S.}~\bibnamefont {Onoda}}, \bibinfo {author} {\bibfnamefont
  {Y.}~\bibnamefont {Onose}}, \bibinfo {author} {\bibfnamefont
  {N.}~\bibnamefont {Nagaosa}}, \ and\ \bibinfo {author} {\bibfnamefont
  {Y.}~\bibnamefont {Tokura}},\ }\href@noop {} {\bibfield  {journal} {\bibinfo
  {journal} {Phys. Rev. Lett.}\ }\textbf {\bibinfo {volume} {99}},\ \bibinfo
  {pages} {086602} (\bibinfo {year} {2007})}\BibitemShut {NoStop}%
\bibitem [{\citenamefont {Xiao}\ \emph {et~al.}(2010)\citenamefont {Xiao},
  \citenamefont {Bauer}, \citenamefont {Uchida}, \citenamefont {Saitoh},
  \citenamefont {Maekawa} \emph {et~al.}}]{xiao2010theory}%
  \BibitemOpen
  \bibfield  {author} {\bibinfo {author} {\bibfnamefont {J.}~\bibnamefont
  {Xiao}}, \bibinfo {author} {\bibfnamefont {G.~E.}\ \bibnamefont {Bauer}},
  \bibinfo {author} {\bibfnamefont {K.-c.}\ \bibnamefont {Uchida}}, \bibinfo
  {author} {\bibfnamefont {E.}~\bibnamefont {Saitoh}}, \bibinfo {author}
  {\bibfnamefont {S.}~\bibnamefont {Maekawa}},  \emph {et~al.},\ }\href@noop {}
  {\bibfield  {journal} {\bibinfo  {journal} {Phys. Rev. B}\ }\textbf {\bibinfo
  {volume} {81}},\ \bibinfo {pages} {214418} (\bibinfo {year}
  {2010})}\BibitemShut {NoStop}%
\bibitem [{\citenamefont {Rezende}\ \emph {et~al.}(2014)\citenamefont
  {Rezende}, \citenamefont {Rodr{\'\i}guez-Su{\'a}rez}, \citenamefont {Cunha},
  \citenamefont {Rodrigues}, \citenamefont {Machado}, \citenamefont {Guerra},
  \citenamefont {Ortiz},\ and\ \citenamefont {Azevedo}}]{rezende2014magnon}%
  \BibitemOpen
  \bibfield  {author} {\bibinfo {author} {\bibfnamefont {S.}~\bibnamefont
  {Rezende}}, \bibinfo {author} {\bibfnamefont {R.}~\bibnamefont
  {Rodr{\'\i}guez-Su{\'a}rez}}, \bibinfo {author} {\bibfnamefont
  {R.}~\bibnamefont {Cunha}}, \bibinfo {author} {\bibfnamefont
  {A.}~\bibnamefont {Rodrigues}}, \bibinfo {author} {\bibfnamefont
  {F.}~\bibnamefont {Machado}}, \bibinfo {author} {\bibfnamefont {G.~F.}\
  \bibnamefont {Guerra}}, \bibinfo {author} {\bibfnamefont {J.~L.}\
  \bibnamefont {Ortiz}}, \ and\ \bibinfo {author} {\bibfnamefont
  {A.}~\bibnamefont {Azevedo}},\ }\href@noop {} {\bibfield  {journal} {\bibinfo
   {journal} {Phys. Rev. B}\ }\textbf {\bibinfo {volume} {89}},\ \bibinfo
  {pages} {014416} (\bibinfo {year} {2014})}\BibitemShut {NoStop}%
\bibitem [{\citenamefont {Arana}\ \emph {et~al.}(2018)\citenamefont {Arana},
  \citenamefont {Gamino}, \citenamefont {Silva}, \citenamefont {Barthem},
  \citenamefont {Givord}, \citenamefont {Azevedo},\ and\ \citenamefont
  {Rezende}}]{arana2018spin}%
  \BibitemOpen
  \bibfield  {author} {\bibinfo {author} {\bibfnamefont {M.}~\bibnamefont
  {Arana}}, \bibinfo {author} {\bibfnamefont {M.}~\bibnamefont {Gamino}},
  \bibinfo {author} {\bibfnamefont {E.}~\bibnamefont {Silva}}, \bibinfo
  {author} {\bibfnamefont {V.}~\bibnamefont {Barthem}}, \bibinfo {author}
  {\bibfnamefont {D.}~\bibnamefont {Givord}}, \bibinfo {author} {\bibfnamefont
  {A.}~\bibnamefont {Azevedo}}, \ and\ \bibinfo {author} {\bibfnamefont
  {S.}~\bibnamefont {Rezende}},\ }\href@noop {} {\bibfield  {journal} {\bibinfo
   {journal} {Phys. Rev. B}\ }\textbf {\bibinfo {volume} {98}},\ \bibinfo
  {pages} {144431} (\bibinfo {year} {2018})}\BibitemShut {NoStop}%
\bibitem [{\citenamefont {Marmion}\ \emph {et~al.}(2014)\citenamefont
  {Marmion}, \citenamefont {Ali}, \citenamefont {McLaren}, \citenamefont
  {Williams},\ and\ \citenamefont {Hickey}}]{marmion2014temperature}%
  \BibitemOpen
  \bibfield  {author} {\bibinfo {author} {\bibfnamefont {S.}~\bibnamefont
  {Marmion}}, \bibinfo {author} {\bibfnamefont {M.}~\bibnamefont {Ali}},
  \bibinfo {author} {\bibfnamefont {M.}~\bibnamefont {McLaren}}, \bibinfo
  {author} {\bibfnamefont {D.}~\bibnamefont {Williams}}, \ and\ \bibinfo
  {author} {\bibfnamefont {B.}~\bibnamefont {Hickey}},\ }\href@noop {}
  {\bibfield  {journal} {\bibinfo  {journal} {Phys. Rev. B}\ }\textbf {\bibinfo
  {volume} {89}},\ \bibinfo {pages} {220404} (\bibinfo {year}
  {2014})}\BibitemShut {NoStop}%
\bibitem [{\citenamefont {Weiler}\ \emph {et~al.}(2013)\citenamefont {Weiler},
  \citenamefont {Althammer}, \citenamefont {Schreier}, \citenamefont {Lotze},
  \citenamefont {Pernpeintner}, \citenamefont {Meyer}, \citenamefont {Huebl},
  \citenamefont {Gross}, \citenamefont {Kamra}, \citenamefont {Xiao} \emph
  {et~al.}}]{weiler2013experimental}%
  \BibitemOpen
  \bibfield  {author} {\bibinfo {author} {\bibfnamefont {M.}~\bibnamefont
  {Weiler}}, \bibinfo {author} {\bibfnamefont {M.}~\bibnamefont {Althammer}},
  \bibinfo {author} {\bibfnamefont {M.}~\bibnamefont {Schreier}}, \bibinfo
  {author} {\bibfnamefont {J.}~\bibnamefont {Lotze}}, \bibinfo {author}
  {\bibfnamefont {M.}~\bibnamefont {Pernpeintner}}, \bibinfo {author}
  {\bibfnamefont {S.}~\bibnamefont {Meyer}}, \bibinfo {author} {\bibfnamefont
  {H.}~\bibnamefont {Huebl}}, \bibinfo {author} {\bibfnamefont
  {R.}~\bibnamefont {Gross}}, \bibinfo {author} {\bibfnamefont
  {A.}~\bibnamefont {Kamra}}, \bibinfo {author} {\bibfnamefont
  {J.}~\bibnamefont {Xiao}},  \emph {et~al.},\ }\href@noop {} {\bibfield
  {journal} {\bibinfo  {journal} {Phys. Rev. Lett.}\ }\textbf {\bibinfo
  {volume} {111}},\ \bibinfo {pages} {176601} (\bibinfo {year}
  {2013})}\BibitemShut {NoStop}%
\bibitem [{\citenamefont {Visser}\ \emph {et~al.}(1997)\citenamefont {Visser},
  \citenamefont {Ramirez},\ and\ \citenamefont
  {Subramanian}}]{visser1997thermal}%
  \BibitemOpen
  \bibfield  {author} {\bibinfo {author} {\bibfnamefont {D.}~\bibnamefont
  {Visser}}, \bibinfo {author} {\bibfnamefont {A.}~\bibnamefont {Ramirez}}, \
  and\ \bibinfo {author} {\bibfnamefont {M.}~\bibnamefont {Subramanian}},\
  }\href@noop {} {\bibfield  {journal} {\bibinfo  {journal} {Phys. Rev Lett.}\
  }\textbf {\bibinfo {volume} {78}},\ \bibinfo {pages} {3947} (\bibinfo {year}
  {1997})}\BibitemShut {NoStop}%
\bibitem [{\citenamefont {Barker}\ and\ \citenamefont
  {Bauer}(2016)}]{barker2016thermal}%
  \BibitemOpen
  \bibfield  {author} {\bibinfo {author} {\bibfnamefont {J.}~\bibnamefont
  {Barker}}\ and\ \bibinfo {author} {\bibfnamefont {G.~E.}\ \bibnamefont
  {Bauer}},\ }\href@noop {} {\bibfield  {journal} {\bibinfo  {journal} {Phys.
  Rev. Lett.}\ }\textbf {\bibinfo {volume} {117}},\ \bibinfo {pages} {217201}
  (\bibinfo {year} {2016})}\BibitemShut {NoStop}%
\bibitem [{\citenamefont {Adachi}\ \emph {et~al.}(2018)\citenamefont {Adachi},
  \citenamefont {Yamamoto},\ and\ \citenamefont {Ichioka}}]{adachi2018spin}%
  \BibitemOpen
  \bibfield  {author} {\bibinfo {author} {\bibfnamefont {H.}~\bibnamefont
  {Adachi}}, \bibinfo {author} {\bibfnamefont {Y.}~\bibnamefont {Yamamoto}}, \
  and\ \bibinfo {author} {\bibfnamefont {M.}~\bibnamefont {Ichioka}},\
  }\href@noop {} {\bibfield  {journal} {\bibinfo  {journal} {J. Phys. D}\
  }\textbf {\bibinfo {volume} {51}},\ \bibinfo {pages} {144001} (\bibinfo
  {year} {2018})}\BibitemShut {NoStop}%
\bibitem [{\citenamefont {Pino}\ \emph {et~al.}(2013)\citenamefont {Pino},
  \citenamefont {Arnache}, \citenamefont {Osorio},\ and\ \citenamefont
  {Tirado}}]{pino2013single}%
  \BibitemOpen
  \bibfield  {author} {\bibinfo {author} {\bibfnamefont {J.}~\bibnamefont
  {Pino}}, \bibinfo {author} {\bibfnamefont {O.}~\bibnamefont {Arnache}},
  \bibinfo {author} {\bibfnamefont {J.}~\bibnamefont {Osorio}}, \ and\ \bibinfo
  {author} {\bibfnamefont {L.}~\bibnamefont {Tirado}},\ }in\ \href@noop {}
  {\emph {\bibinfo {booktitle} {J. Phys. Conf. Ser.}}},\ Vol.\ \bibinfo
  {volume} {466}\ (\bibinfo {organization} {IOP Publishing},\ \bibinfo {year}
  {2013})\ p.\ \bibinfo {pages} {012021}\BibitemShut {NoStop}%
\bibitem [{\citenamefont {Huhtinen}\ \emph
  {et~al.}(2002{\natexlab{b}})\citenamefont {Huhtinen}, \citenamefont {Laiho},
  \citenamefont {L{\"a}hderanta}, \citenamefont {Salminen}, \citenamefont
  {Lisunov},\ and\ \citenamefont {Zakhvalinskii}}]{huhtinen2002unconventional}%
  \BibitemOpen
  \bibfield  {author} {\bibinfo {author} {\bibfnamefont {H.}~\bibnamefont
  {Huhtinen}}, \bibinfo {author} {\bibfnamefont {R.}~\bibnamefont {Laiho}},
  \bibinfo {author} {\bibfnamefont {E.}~\bibnamefont {L{\"a}hderanta}},
  \bibinfo {author} {\bibfnamefont {J.}~\bibnamefont {Salminen}}, \bibinfo
  {author} {\bibfnamefont {K.}~\bibnamefont {Lisunov}}, \ and\ \bibinfo
  {author} {\bibfnamefont {V.}~\bibnamefont {Zakhvalinskii}},\ }\href@noop {}
  {\bibfield  {journal} {\bibinfo  {journal} {J. Appl. Phys.}\ }\textbf
  {\bibinfo {volume} {91}},\ \bibinfo {pages} {7944} (\bibinfo {year}
  {2002}{\natexlab{b}})}\BibitemShut {NoStop}%
\bibitem [{\citenamefont {Kim}\ \emph {et~al.}(2002)\citenamefont {Kim},
  \citenamefont {Revaz}, \citenamefont {Zink}, \citenamefont {Hellman},
  \citenamefont {Rhyne},\ and\ \citenamefont {Mitchell}}]{kim2002tricritical}%
  \BibitemOpen
  \bibfield  {author} {\bibinfo {author} {\bibfnamefont {D.}~\bibnamefont
  {Kim}}, \bibinfo {author} {\bibfnamefont {B.}~\bibnamefont {Revaz}}, \bibinfo
  {author} {\bibfnamefont {B.}~\bibnamefont {Zink}}, \bibinfo {author}
  {\bibfnamefont {F.}~\bibnamefont {Hellman}}, \bibinfo {author} {\bibfnamefont
  {J.}~\bibnamefont {Rhyne}}, \ and\ \bibinfo {author} {\bibfnamefont
  {J.}~\bibnamefont {Mitchell}},\ }\href@noop {} {\bibfield  {journal}
  {\bibinfo  {journal} {Phys. Rev. Lett.}\ }\textbf {\bibinfo {volume} {89}},\
  \bibinfo {pages} {227202} (\bibinfo {year} {2002})}\BibitemShut {NoStop}%
\bibitem [{\citenamefont {Isasa}\ \emph {et~al.}(2015)\citenamefont {Isasa},
  \citenamefont {Villamor}, \citenamefont {Hueso}, \citenamefont {Gradhand},\
  and\ \citenamefont {Casanova}}]{isasa2015temperature}%
  \BibitemOpen
  \bibfield  {author} {\bibinfo {author} {\bibfnamefont {M.}~\bibnamefont
  {Isasa}}, \bibinfo {author} {\bibfnamefont {E.}~\bibnamefont {Villamor}},
  \bibinfo {author} {\bibfnamefont {L.~E.}\ \bibnamefont {Hueso}}, \bibinfo
  {author} {\bibfnamefont {M.}~\bibnamefont {Gradhand}}, \ and\ \bibinfo
  {author} {\bibfnamefont {F.}~\bibnamefont {Casanova}},\ }\href@noop {}
  {\bibfield  {journal} {\bibinfo  {journal} {Phys. Rev. B}\ }\textbf {\bibinfo
  {volume} {91}},\ \bibinfo {pages} {024402} (\bibinfo {year}
  {2015})}\BibitemShut {NoStop}%
\bibitem [{\citenamefont {Lynn}\ \emph {et~al.}(1996)\citenamefont {Lynn},
  \citenamefont {Erwin}, \citenamefont {Borchers}, \citenamefont {Huang},
  \citenamefont {Santoro}, \citenamefont {Peng},\ and\ \citenamefont
  {Li}}]{lynn1996unconventional}%
  \BibitemOpen
  \bibfield  {author} {\bibinfo {author} {\bibfnamefont {J.}~\bibnamefont
  {Lynn}}, \bibinfo {author} {\bibfnamefont {R.}~\bibnamefont {Erwin}},
  \bibinfo {author} {\bibfnamefont {J.}~\bibnamefont {Borchers}}, \bibinfo
  {author} {\bibfnamefont {Q.}~\bibnamefont {Huang}}, \bibinfo {author}
  {\bibfnamefont {A.}~\bibnamefont {Santoro}}, \bibinfo {author} {\bibfnamefont
  {J.}~\bibnamefont {Peng}}, \ and\ \bibinfo {author} {\bibfnamefont
  {Z.}~\bibnamefont {Li}},\ }\href@noop {} {\bibfield  {journal} {\bibinfo
  {journal} {Phys. Rev. Lett.}\ }\textbf {\bibinfo {volume} {76}},\ \bibinfo
  {pages} {4046} (\bibinfo {year} {1996})}\BibitemShut {NoStop}%
\bibitem [{\citenamefont {Jermain}\ \emph {et~al.}(2017)\citenamefont
  {Jermain}, \citenamefont {Aradhya}, \citenamefont {Reynolds}, \citenamefont
  {Buhrman}, \citenamefont {Brangham}, \citenamefont {Page}, \citenamefont
  {Hammel}, \citenamefont {Yang},\ and\ \citenamefont
  {Ralph}}]{jermain2017increased}%
  \BibitemOpen
  \bibfield  {author} {\bibinfo {author} {\bibfnamefont {C.}~\bibnamefont
  {Jermain}}, \bibinfo {author} {\bibfnamefont {S.}~\bibnamefont {Aradhya}},
  \bibinfo {author} {\bibfnamefont {N.}~\bibnamefont {Reynolds}}, \bibinfo
  {author} {\bibfnamefont {R.}~\bibnamefont {Buhrman}}, \bibinfo {author}
  {\bibfnamefont {J.}~\bibnamefont {Brangham}}, \bibinfo {author}
  {\bibfnamefont {M.}~\bibnamefont {Page}}, \bibinfo {author} {\bibfnamefont
  {P.}~\bibnamefont {Hammel}}, \bibinfo {author} {\bibfnamefont
  {F.}~\bibnamefont {Yang}}, \ and\ \bibinfo {author} {\bibfnamefont
  {D.}~\bibnamefont {Ralph}},\ }\href@noop {} {\bibfield  {journal} {\bibinfo
  {journal} {Phys. Rev. B}\ }\textbf {\bibinfo {volume} {95}},\ \bibinfo
  {pages} {174411} (\bibinfo {year} {2017})}\BibitemShut {NoStop}%
\bibitem [{\citenamefont {Maier-Flaig}\ \emph {et~al.}(2017)\citenamefont
  {Maier-Flaig}, \citenamefont {Klingler}, \citenamefont {Dubs}, \citenamefont
  {Surzhenko}, \citenamefont {Gross}, \citenamefont {Weiler}, \citenamefont
  {Huebl},\ and\ \citenamefont {Goennenwein}}]{maier2017temperature}%
  \BibitemOpen
  \bibfield  {author} {\bibinfo {author} {\bibfnamefont {H.}~\bibnamefont
  {Maier-Flaig}}, \bibinfo {author} {\bibfnamefont {S.}~\bibnamefont
  {Klingler}}, \bibinfo {author} {\bibfnamefont {C.}~\bibnamefont {Dubs}},
  \bibinfo {author} {\bibfnamefont {O.}~\bibnamefont {Surzhenko}}, \bibinfo
  {author} {\bibfnamefont {R.}~\bibnamefont {Gross}}, \bibinfo {author}
  {\bibfnamefont {M.}~\bibnamefont {Weiler}}, \bibinfo {author} {\bibfnamefont
  {H.}~\bibnamefont {Huebl}}, \ and\ \bibinfo {author} {\bibfnamefont {S.~T.}\
  \bibnamefont {Goennenwein}},\ }\href@noop {} {\bibfield  {journal} {\bibinfo
  {journal} {Phys. Rev. B}\ }\textbf {\bibinfo {volume} {95}},\ \bibinfo
  {pages} {214423} (\bibinfo {year} {2017})}\BibitemShut {NoStop}%
\bibitem [{\citenamefont {Heinrich}\ \emph {et~al.}(2011)\citenamefont
  {Heinrich}, \citenamefont {Burrowes}, \citenamefont {Montoya}, \citenamefont
  {Kardasz}, \citenamefont {Girt}, \citenamefont {Song}, \citenamefont {Sun},\
  and\ \citenamefont {Wu}}]{heinrich2011spin}%
  \BibitemOpen
  \bibfield  {author} {\bibinfo {author} {\bibfnamefont {B.}~\bibnamefont
  {Heinrich}}, \bibinfo {author} {\bibfnamefont {C.}~\bibnamefont {Burrowes}},
  \bibinfo {author} {\bibfnamefont {E.}~\bibnamefont {Montoya}}, \bibinfo
  {author} {\bibfnamefont {B.}~\bibnamefont {Kardasz}}, \bibinfo {author}
  {\bibfnamefont {E.}~\bibnamefont {Girt}}, \bibinfo {author} {\bibfnamefont
  {Y.-Y.}\ \bibnamefont {Song}}, \bibinfo {author} {\bibfnamefont
  {Y.}~\bibnamefont {Sun}}, \ and\ \bibinfo {author} {\bibfnamefont
  {M.}~\bibnamefont {Wu}},\ }\href@noop {} {\bibfield  {journal} {\bibinfo
  {journal} {Phys. Rev. Lett.}\ }\textbf {\bibinfo {volume} {107}},\ \bibinfo
  {pages} {066604} (\bibinfo {year} {2011})}\BibitemShut {NoStop}%
\bibitem [{\citenamefont {Mosendz}\ \emph {et~al.}(2010)\citenamefont
  {Mosendz}, \citenamefont {Pearson}, \citenamefont {Fradin}, \citenamefont
  {Bauer}, \citenamefont {Bader},\ and\ \citenamefont
  {Hoffmann}}]{mosendz2010quantifying}%
  \BibitemOpen
  \bibfield  {author} {\bibinfo {author} {\bibfnamefont {O.}~\bibnamefont
  {Mosendz}}, \bibinfo {author} {\bibfnamefont {J.}~\bibnamefont {Pearson}},
  \bibinfo {author} {\bibfnamefont {F.}~\bibnamefont {Fradin}}, \bibinfo
  {author} {\bibfnamefont {G.}~\bibnamefont {Bauer}}, \bibinfo {author}
  {\bibfnamefont {S.}~\bibnamefont {Bader}}, \ and\ \bibinfo {author}
  {\bibfnamefont {A.}~\bibnamefont {Hoffmann}},\ }\href@noop {} {\bibfield
  {journal} {\bibinfo  {journal} {Phys. Rev. Lett.}\ }\textbf {\bibinfo
  {volume} {104}},\ \bibinfo {pages} {046601} (\bibinfo {year}
  {2010})}\BibitemShut {NoStop}%
\bibitem [{\citenamefont {Nakayama}\ \emph {et~al.}(2012)\citenamefont
  {Nakayama}, \citenamefont {Ando}, \citenamefont {Harii}, \citenamefont
  {Yoshino}, \citenamefont {Takahashi}, \citenamefont {Kajiwara}, \citenamefont
  {Uchida}, \citenamefont {Fujikawa},\ and\ \citenamefont
  {Saitoh}}]{nakayama2012geometry}%
  \BibitemOpen
  \bibfield  {author} {\bibinfo {author} {\bibfnamefont {H.}~\bibnamefont
  {Nakayama}}, \bibinfo {author} {\bibfnamefont {K.}~\bibnamefont {Ando}},
  \bibinfo {author} {\bibfnamefont {K.}~\bibnamefont {Harii}}, \bibinfo
  {author} {\bibfnamefont {T.}~\bibnamefont {Yoshino}}, \bibinfo {author}
  {\bibfnamefont {R.}~\bibnamefont {Takahashi}}, \bibinfo {author}
  {\bibfnamefont {Y.}~\bibnamefont {Kajiwara}}, \bibinfo {author}
  {\bibfnamefont {K.-i.}\ \bibnamefont {Uchida}}, \bibinfo {author}
  {\bibfnamefont {Y.}~\bibnamefont {Fujikawa}}, \ and\ \bibinfo {author}
  {\bibfnamefont {E.}~\bibnamefont {Saitoh}},\ }\href@noop {} {\bibfield
  {journal} {\bibinfo  {journal} {Phys. Rev. B}\ }\textbf {\bibinfo {volume}
  {85}},\ \bibinfo {pages} {144408} (\bibinfo {year} {2012})}\BibitemShut
  {NoStop}%
\bibitem [{\citenamefont {Dyakonov}\ \emph {et~al.}(2003)\citenamefont
  {Dyakonov}, \citenamefont {Shapovalov}, \citenamefont {Zubov}, \citenamefont
  {Aleshkevych}, \citenamefont {Klimov}, \citenamefont {Varyukhin},
  \citenamefont {Pashchenko}, \citenamefont {Kamenev}, \citenamefont
  {Mikhailov}, \citenamefont {Dyakonov} \emph
  {et~al.}}]{dyakonov2003ferromagnetic}%
  \BibitemOpen
  \bibfield  {author} {\bibinfo {author} {\bibfnamefont {V.}~\bibnamefont
  {Dyakonov}}, \bibinfo {author} {\bibfnamefont {V.}~\bibnamefont
  {Shapovalov}}, \bibinfo {author} {\bibfnamefont {E.}~\bibnamefont {Zubov}},
  \bibinfo {author} {\bibfnamefont {P.}~\bibnamefont {Aleshkevych}}, \bibinfo
  {author} {\bibfnamefont {A.}~\bibnamefont {Klimov}}, \bibinfo {author}
  {\bibfnamefont {V.}~\bibnamefont {Varyukhin}}, \bibinfo {author}
  {\bibfnamefont {V.}~\bibnamefont {Pashchenko}}, \bibinfo {author}
  {\bibfnamefont {V.}~\bibnamefont {Kamenev}}, \bibinfo {author} {\bibfnamefont
  {V.}~\bibnamefont {Mikhailov}}, \bibinfo {author} {\bibfnamefont
  {K.}~\bibnamefont {Dyakonov}},  \emph {et~al.},\ }\href@noop {} {\bibfield
  {journal} {\bibinfo  {journal} {J. Appl. Phys.}\ }\textbf {\bibinfo {volume}
  {93}},\ \bibinfo {pages} {2100} (\bibinfo {year} {2003})}\BibitemShut
  {NoStop}%
\end{thebibliography}%

\end{document}